\documentclass[aps,prl,twocolumn,groupedaddress]{revtex4}

\usepackage{graphicx}
\usepackage{dcolumn}
\usepackage{bm}

\def\be{\begin{equation}}
\def\ee{\end{equation}}
\def\bea{\begin{eqnarray}}
\def\eea{\end{eqnarray}}

\newcommand{\YBCO}{YBa$_2$Cu$_3$O$_{x}$}

\newcommand{\LSCO}{La$_{2-x}$Sr$_x$CuO$_{4}$}

\begin{document}


\title{Experimental investigation of  the magnetic field driven superconductor/ insulator transition in underdoped \LSCO\  thin films}
\author{Brigitte Leridon}
\affiliation{Laboratoire de Physique et  d'Etude des Mat\'eriaux,  UMR8213/CNRS -  ESPCI ParisTech - UPMC, 10 rue Vauquelin, 75005 Paris, France}
\author{J. Vanacken and V.V. Moshchalkov}
\affiliation{
Institute for Nanoscale Physics and Chemistry, Katholieke Universiteit Leuven, Celestijnenlaan 200 D, B-3001 Heverlee, Belgium }
\author{Baptiste Vignolle}
\affiliation{Laboratoire National des Champs Magn\'etiques Intenses, 
143 Avenue de Rangueil, 31400 Toulouse, France}

\author{Rajni Porwal and R.C. Budhani}
\affiliation{National Physical Laboratory, New Delhi  110012 and, Indian Institute of Technology, Kanpur, India}

\begin{abstract}

The magnetic field driven superconductor/insulator transition is studied in a large variety of \LSCO\ thin films of various Sr dopings. Temperature dependence of the resistivity down to 4.2 or 1.5 K  under high pulsed magnetic field (up to 57 T) is analyzed.  In particular, the existence of plateaus in the resistance versus temperature curves, in a limited range of temperature, for given values of the magnetic field is carefully investigated. It is shown to be associated to scaling behaviour of the resistance  versus magnetic field curves, evocative of the presence of a quantum critical point. 
A  three-dimensional (H,x,T) phase diagram is proposed, taking into account the intrinsic lamellar nature of the materials by the existence of a temperature crossover from quantum-two-dimensional to three-dimensional behavior. 

\end{abstract}

\maketitle

\section{introduction}

Theory predicts that the resistance of a two-dimensional (2D) electron gas at zero temperature should be either zero or infinite \cite{Lee:1985}.  Due to the fact that a small amount of disorder is able to localize all the fermions states, the only possibility is an insulator, unless non-zero pairing interactions are present and yield superconductivity.  It has therefore been proposed that  a disordered 2D electron system could undergo a quantum superconductor/insulator transition (QSIT) driven either by disorder or by magnetic field \cite{Fisher:1990b, Fisher:1990a}.    The mechanism for this transition is that the  vortex/antivortex pairs that are non mobile below a certain disorder threshold, may transform into a Bose condensate of free vortices above this threshold, while the Cooper pairs localize, leading to the formation of a Bose insulator.  When the transition is driven by a magnetic field, instead, the system goes from a vortex glass state to a Bose insulator. Consequently, there should be only one value of the magnetic field $H=H_C$ or of disorder $\delta=\delta_C$ at which the system at zero temperature is metallic. This corresponds to the observation of a plateau in the resistance versus temperature curve for H$_C$ or $\delta_C$.  The corresponding value of the resistance per square $R_C$ at the plateau was originally predicted to be  equal to $h/4e^2$ (in the case of a self-dual model)\cite{Fisher:1990b,Fisher:1990a}. However, more recent theoretical work taking into account a two-fluid model show that a resistive phase may interpolate between the superconducting and insulating phases \cite{Galitski:2005}  and make the resistance at the transition deviate from this universal value.

Evidence has been found of such a quantum transition in granular films of In/InOx by varying the composition  \cite{Hebard:1984} or the magnetic field \cite{Hebard:1990}, but also in homogeneous ultrathin films of Bi and Pb \cite{Haviland:1989} by varying the layer thickness (which is considered equivalent as varying the disorder), and more recently in MoGe \cite{Graybeal:1984,Yazdani:1995}and  NbSi \cite{Aubin:2006}.  In most of the experiments, the films were seen to undergo a transition towards insulating state for a resistance per square only of the order of  (and not exactly) $h/4e^2$.

Since high-T$_C$ cuprate superconductors are known to be intrinsically lamellar, it has been proposed \cite{Steiner:2005} that a similar  quantum transition is at play also in such systems where the system undergoes a field-tuned superconductor to insulator transition at low temperature.  The disorder-tuned transition had been studied in these systems by Wang and coworkers \cite{Wang:1991}.

Recently  Bollinger and coworkers \cite{Bollinger:2011} have reported superconducting/insulator transition driven by an\textit{ electrical field} in an  extremely thin layer (one or two unit cells) of optimally doped  \LSCO .  In this case the transition is observed for a carrier concentration of p=0.06. The number of carriers is varied through application of a gate voltage and data of resistance versus temperature is collected for different gate voltages.  
Another group has reported similar measurements on \YBCO   thin films for a carrier concentration estimated to be around $x=0.05$ \cite{Leng:2011}.

We will investigate here the behavior of resistance versus temperature of typically 100nm-thick  \LSCO\  thin films of different Sr concentrations while varying the magnetic field and the temperature.  In particular, we will carefully inspect the possibility of existence of a quantum  superconducting-insulator transition (QSIT) in those systems.

\section{Experiments}

The as-grown La$_{2-x}$Sr$_x$CuO$_4$ films were prepared by DC magnetron sputtering from stoechiometric targets either at KU LeuvenÊ\cite{Wagner:2001} or by pulsed laser deposition at ITT Kanpur (see the elaboration conditions in  \cite{Rout:2010}).Ê The critical temperatures were determined  as the  transition mid-point temperatures obtained by measuring the resistance under zero magnetic field.

Part of the transport measurements were carried out at the KU Leuven high field facility and the other part at the Toulouse LNCMI high field facility.
Pulsed high-field measurements up to 49-57~T were performed from 4.2~K to 300~K on nine \LSCO thin films with different Sr concentration x. The c-axis orientated films were mounted in general with $\mu_0H // c$ (perpendicular field),Ê and the current (of typically mA and 50 kHz) was along the ab-plane (IÊ//Êab).Ê 
Among the different samples, some were measured down to 1.5~K and up to 58~T and one  was measured both in parallel and perpendicular magnetic field. 

\subsection{High-field resistivity as a function of temperature}

At maximal field (between 49~T and 57~T according to the different experiments),  the resistance of the underdoped samples versus temperature varies approximately as $Ln(1/T)$ as was reported previously \cite{Ando:1995,Leridon:2007b}, which is consistent in particular with weak localization of vortices as proposed by Das and Doniach \cite{Das:1999,Das:2001}.  It is noteworthy that for the underdoped samples that are \textit{not} superconducting at zero magnetic field (i.e. $x<0.06$), the resistance versus temperature at zero magnetic field exhibits a Shklovskii-Efros localization law ($\rho=\rho_0 exp((T/T_0)^{1/2}$) instead of a $Ln(1/T)$, which seems to indicate that the nature of the insulating state obtained from a superconductor under  50~T may be different from the nature of the insulating state at low doping levels. (See the data in Weckuysen et al. \cite{Weckuysen:2002b} ).
It may more likely be an indication that $\rho=\rho_0 Ln(1/T)$, whatever its physical origin,  probably related to granular metallicity,  is the strongest insulating behavior allowing superconductivity to develop \footnote{M. Grilli. Private communication}. This is strongly supported by measurements in inhomogeneous conventional superconductors, where such a logarithmic behavior is observed up to 300K  in NbN cermets when superconductivity develops at  low temperature and strong-localization conductivity is observed whenever the samples are not superconducting \cite{Deutscher:1980,Simon:1987}.

\begin{figure}
\includegraphics[width=0.5\textwidth]{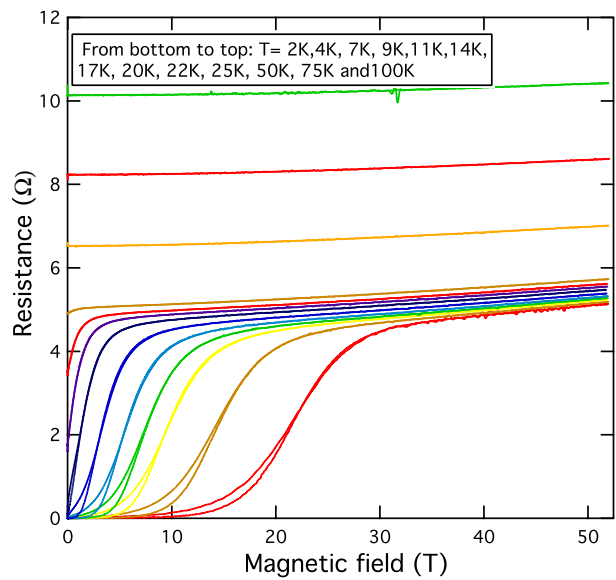}
\caption{(Color on line.) Resistance versus perpendicular magnetic field data for an overdoped \LSCO\ thin film corresponding to $x=0.25$ and $T_C$=18.2~K. 
 The absence of crossing between the R(H) curves taken at different temperatures means that there is no plateau in the R(T) curves.}
\label{overdoped}
\end{figure}

\subsection{Behaviour of overdoped samples  ($x>0.2$)}

On overdoped samples ($x>0.2$) the R(H) curves measured in perpendicular magnetic field show no crossing point since no upturn of the resistance versus temperature is present at low temperatures for fields as high as 50~T.  For this range of doping levels the system is metallic even down to low temperature, rather than insulating.
(See for example the resistance versus magnetic field data in Figure \ref{overdoped} where no crossing point is present at all.)
We shall now focus on the underdoped and optimally doped samples.

\begin{figure}
\includegraphics[width=0.5\textwidth]{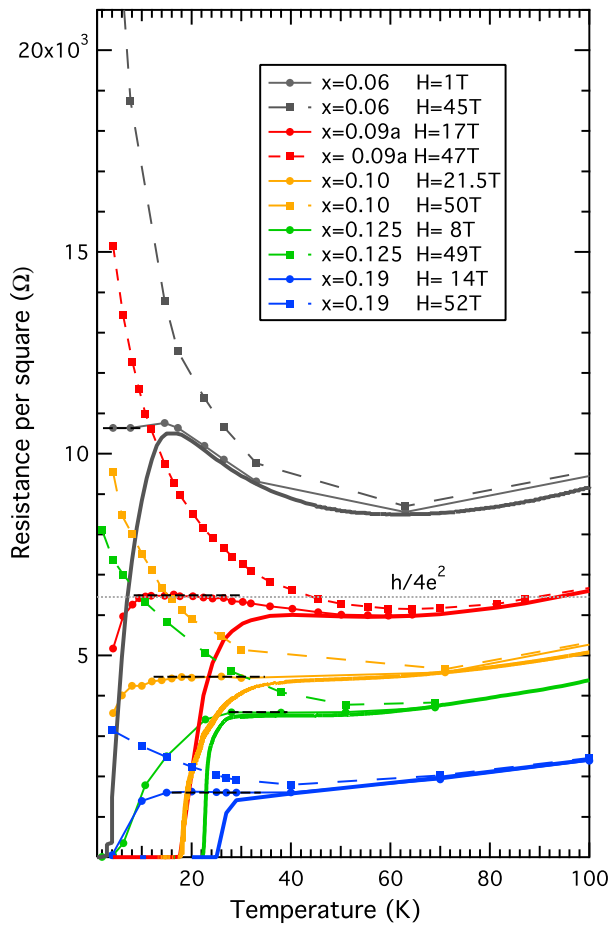}
\caption{Resistance per square of CuO$_2$ layer versus temperature for five representative samples. Solid lines H=0T; circles and lines:  H=H$_C$ (the values of H$_C$ are indicated in the legend); squares and dashed lines:  square resistance  at maximal field, as indicated in the legend. The plateaus are highlighted with dashed thick lines. The thin grey dotted line denotes the position of the resistance quantum $h/4e^2$.}
\label{plat1}
\end{figure}

\begin{figure}
\includegraphics[width=0.5\textwidth]{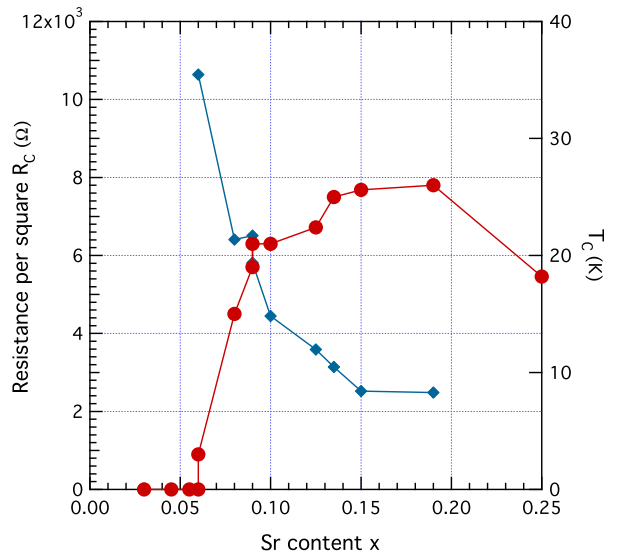}
\caption{(Color on line.) Critical temperature $T_C$ (red dots, right scale) and square resistance $R_C$ (blue squares, left scale) at the plateau as a function of the Sr content.}
\label{Rcarreplat}
\end{figure}

\begin{figure}
\includegraphics[width=0.5\textwidth]{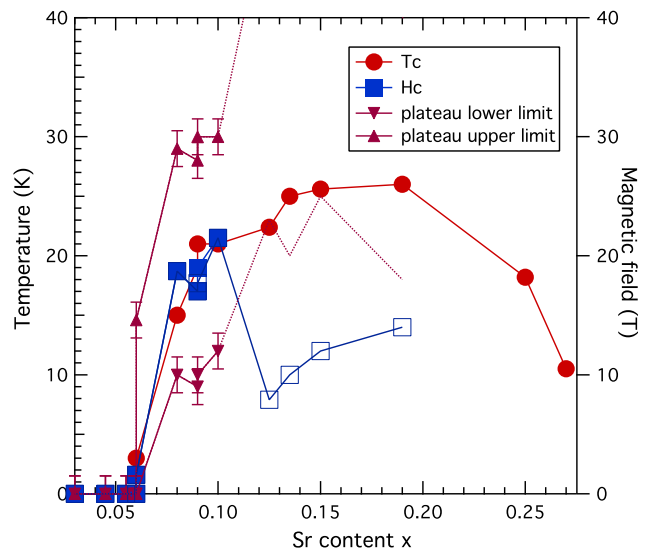}
\caption{(Color on line.) Critical temperature $T_C$ (red dots, left scale) and critical field $H_C$ for which a plateau in the R(T) curve is observed (blue squares, right scale), as a function of the Sr content $x$. Note that for  $x\leq0.1$, the scales for $H_C$ and $T_C$ are the same ($1T\approx1K$). The variation of $H_C$ displays an abrupt change at about $x=0.125$, pointing toward different origins for the plateaux below and above $x=0.125$. The down and up pink triangles (and dotted pink line) denote the respective lower and upper temperature limits for the observation of the plateaus, defined for a spread in the resistance values of less than 1\%. The error bars on the triangles correspond to the typical temperature spacing. The open squares (as opposed to filled squares) indicate a different origin for the plateaus for $x\geq0.125$ since no scaling is associated (see text).}
\label{phadiag}
\end{figure}

\subsection{Underdoped samples and optimally doped samples $x\leq0.19$}

In previous work from Ando and coworkers \cite{Ando:1995}, a negative magnetoresistance is shown at high magnetic field. This was argued to plead in favor of the presence of a Bose insulator by \cite{Steiner:2005}. The maximum of the resistance versus field was hypothesized to be H$_C^2$ and the negative magnetoresistance was attributed to localized-pair breaking. We did not observed such behaviour here in samples with similar Sr concentrations and in the same field range. Although the measured voltage did exhibit such a negative slope, when calculating the resistance by dividing by the effective current through the sample that was measured during the pulse, the negative MR was not present. 

By contrast to the overdoped case, on every underdoped and optimally doped sample, a fixed crossing point in the $R(H)$ data was observed \textit{within a finite range of temperatures} for $H=H_C(x)$. (This crossing point was seen in a range of temperature where the data in ref \cite{Ando:1995} are not shown.) This leads to a plateau in the R versus T curves when the magnetic field is equal to $H_C$. 

In Figure \ref{plat1} are displayed the resistances per square of CuO$_2$ plane  versus temperature  for a selected set of five representative samples with x ranging from 0.06 to 0.19, for the field values $H=H_C$ (circles and lines) and $H=0~T$ (solid lines).  The resistance at maximal field is also shown (squares and dashed lines). The resistance at the plateau is depicted in Figure \ref{Rcarreplat}  as a function of the Sr content and varies monotonically. (However the two samples LSCO$_{0.009a}$ and  LSCO$_{0.009b}$ of different origins show some small discrepancies, probably due to different level of disorder.)

Figure \ref{phadiag} summarizes the values of $T_C$ taken at the transition inflexion point (reds dots), $H_C$ (blue squares) and the temperature range of observation of the plateaus for different LSCO thin films as a function of the Sr content x.   The temperature range of observation of the plateau, denoted by up and down pink triangles  for $x\leq0.100$ and by a pink dotted line for x$\geq0.125$, corresponds to a spread in the resistance values of less than $1\%$. The error bars on the triangles correspond to the typical temperature spacing between different measurements. 

As may be seen from both Figures \ref{plat1} and \ref{phadiag}, there is a great discrepancy between the data for $x<0.125$ and $x\geq0.125$.  $H_C$ increases with doping and follows the variation of $T_C$ for $x \leq0.100$,  then decreases abruptly at $x=0.125$ and then starts increasing again. As a matter of fact, $H_C$ is found to scale with $T_C$ for $x\leq0.100$ with 1K=1T.
 Although for all  $x\leq0.100$ the plateau develops at temperatures well below the superconductivity onset temperature at zero field, on the contrary for  $x\simeq0.125$ it is only present at temperatures above the foot of the transition at H=0T. 
Besides, for $x\leq0.100$ the resistance at the plateau is higher than the minimum of the full field resistivity curve and for $x\geq0.125$ it is slightly lower. Moreover, for samples with x$\geq0.125$, our attempt to obtain  scaling of the data around the apparent crossing point - as will be described in the following paragraph, revealed to be unsuccessful.

These four observations point towards different explanations for the existence of the plateaus for x lower or greater than $1/8$.  For the following, we will focus on the plateau observable for $x\leq0.1$.

\subsection{Scaling analysis of the resistance for $x<0.125$}

\begin{figure}
\includegraphics[width=0.5\textwidth]{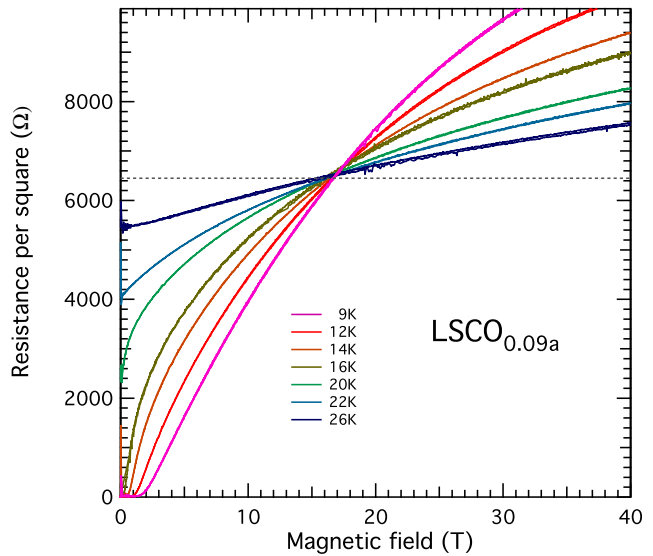}
\caption{Color on line. R(H) data for sample LSCO$_{0.09a}$($x=0.09$) for temperatures between 9K (lower curve at low fields)and 26 K (upper curve at low fields). Note the position of the resistivity quantum $h/4e{^2}$ (grey dotted line).  }
\label{RH644}
\end{figure}

\begin{figure}
\includegraphics[width=0.5\textwidth]{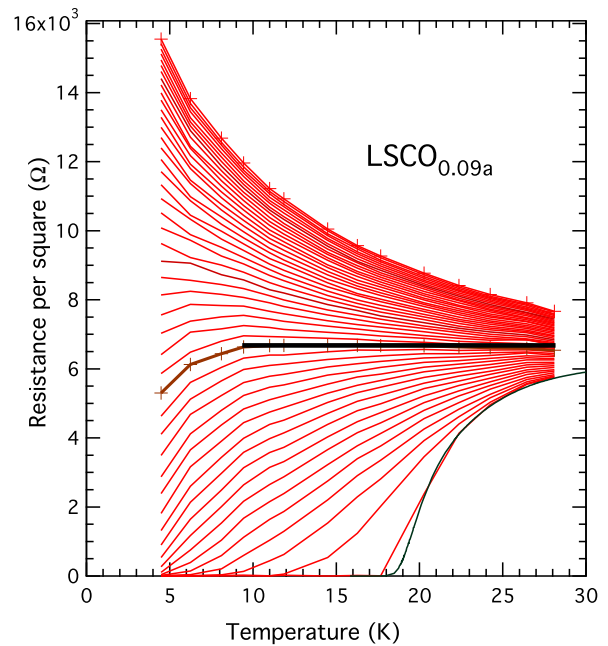}
\caption{(Color on line). R(T) data for different magnetic field values, ranging from 0~T to 47~T by steps of 1~T for sample LSCO$_{0.09a}$ ($x=0.09$) (same as Fig. \ref{RH644}). The brown line, showing a plateau from about 9~K to about 28~K (marked in black), corresponds to 17~T.   The resistance per square at the plateau is about $h/4e^2$.}
\label{RT644}
\end{figure}

\begin{figure}
\includegraphics[width=0.5\textwidth]{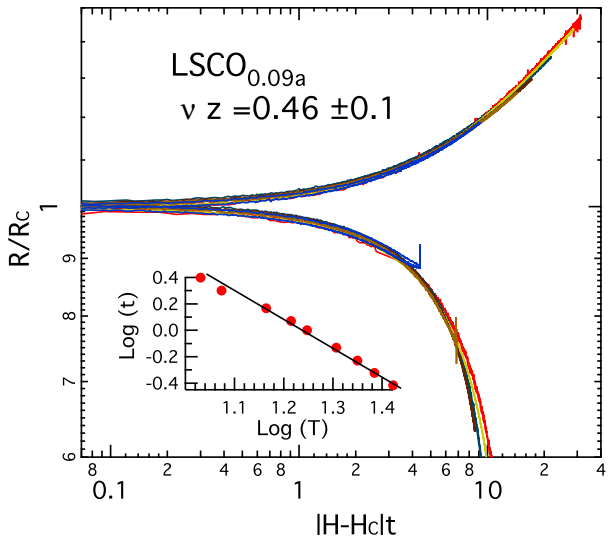}
\caption{Scaling of the data for sample LSCO$_{0.09a}$, $R/R_C=f(|H-H_C|t)$ , with $t=T^{-1/\nu z}$ and $\nu z=0.46\pm0.1$
 }
\label{scal}
\end{figure}

Figure \ref{RH644} shows typical R(H) data curves for one of the samples whose Sr content is x=0.09 (LSCO$_{0.09a}$); a fixed point is observed for temperatures between  9~K and  26~K for $R_C=6.6 k\Omega,H_f=17~T$. This value of resistance is remarquably close to the resistance per square valid for a self dual transition $R=h/{4e^2}$, although the temperature plateau is expected in this case to extend to zero temperature.

 \begin{table}
 \label{T1}
 \begin{center}
 \caption{Summary of the scaling properties. }
\begin{tabular} {c | c | c | c | c | c }

 \\
Sample  & Sr & $T_{C} $ &H$_C$&  $R_C$ &  $ \nu z $\\ 
name & content & K  & T & Ohms &      \\
 \hline
 & &  &  &  & \\
LSCO0.06& 0.06&3&1.6&10638 & -\\
LSCO0.08 &  0.08& 15 & 18.7& 6408 &0.54 $\pm $0.1\\
LSCO0.09a & 0.09 & 19 &16.8 &   6509 &  0.46 $ \pm$ 0.1\\
LSCO0.09b & 0.09 &21 &19 &  5811 &  0.43 $ \pm$ 0.1\\
LSCO0.10 & 0.10 & 21 &21.5 &  4445  &   0.63 $\pm$ 0.1 \\

\end{tabular}
 \end{center}
 \end{table}

The resistance per square of CuO plane versus temperature values are plotted in Fig. \ref{RT644} for sample LSCO${0.09a}$. A plateau is clearly visible for H=17~T, which corresponds to a resistivity of about $h/4e^2$. The departure from the plateau at low temperature is also visible. This figure is typical of what was observed in all the samples with x ranging from 0.06 to 0.1.
However,we believe that the fact that the square resistance is in this case almost equal to $h/4e^2$ at the plateau is rather fortuitous since R$_C$ is found to vary with doping, as may be seen in Table~1  or in Fig. \ref{plat1}. 

Scaling analysis was performed at the vicinity of the fixed point. Figure \ref{scal} shows the scaling of the same curves for sample LSCO$_{0.009a}$, as $R/{R_C}=f(|H-H_C|T^{-1/\nu z})$ with $\nu z = 0.46\pm0.1$.  The scaling exponent $\nu z$ - summarized in Table 1- is found to vary from 0.43$\pm 0.1$  to $0.63\pm0.1$ for the four different samples on which scaling was possible.  It is therefore possible to infer the values of $\nu$, assuming  z=1. These values are lower than the exponent predicted for a (2D+1) quantum xy model ($\nu$z=2/3) and found in amorphous Bi (\cite{Parendo:2005}), or $NbSi$ compounds \cite{Aubin:2006}. They are much lower than the exponent predicted in the framework of a dirty boson picture ($\nu >1$), or  for classical percolation ($\nu=1.3$)  and than the exponent observed for the electrical field driven transition \cite{Bollinger:2011}  ($\nu z =3/2$).

For comparison we also attempted to scale the R(H) curves for the compounds with $x\geq0.125$, which revealed unsuccessful.

\subsection{Further experiments}

\begin{figure}
\includegraphics[width=0.5\textwidth]{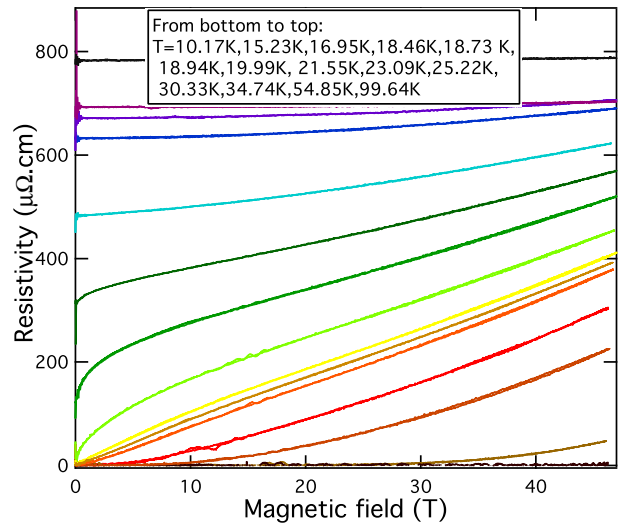}
\caption{R(H) data for all temperatures in parallel magnetic fields for sample LSCO$_{0.09a}$ ($x=0.09$). The absence of a crossing point in the range of measurements is an indication that the mechanism for the transition is quite different from in the perpendicular field experiment. }
\label{parallel}
\end{figure}

Although one important criterium for the observation of a magnetically driven QSIT is the observation of a plateau in the resistance versus temperature data in perpendicular magnetic field,  on the contrary when the magnetic field is applied along the plane of  the layer then the transition should be classically driven and no crossing point should be observed.
As a further test, we measured sample LSCO$_{0.09a}$ ($x=0.09$) again with the magnetic field  applied along the planes, and by contrast to the perpendicular case, the $R(H)$ curves do no longer show a crossing point in our range of measurement as can be seen in Figure \ref{parallel}.  The transition is governed by a different mechanism in this case. In the perpendicular case, the transition is controlled by the binding and debinding of vortices (or vortices dislocations). However,  in the parallel case, since vortices can not form, the transition point is reached only for a much higher (not attainable) value of the magnetic field, corresponding to the breaking of the Cooper pairs.

\begin{figure}
\includegraphics[width=0.5\textwidth]{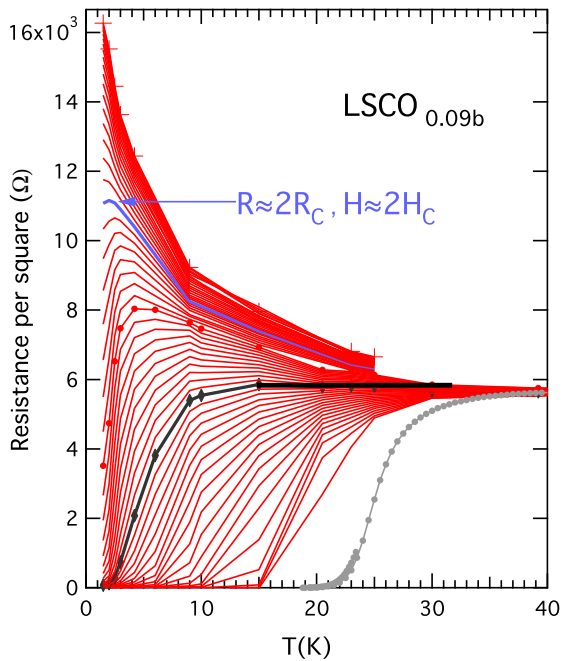}
\caption{(Color on line). R(T) data for sample LSCO0.09b (x=0.09) for different magnetic field values, ranging from 0~T to 56~T by steps of 1~T. The dark grey curve, exhibiting a plateau (marked in black) from less than 15~K up to about 30~K, corresponds to $H_C$=19~T.  The zero temperature SIT occurs for (H$\approx 37T, R\approx11k\Omega$). See the blue curve designated by the arrow.  The resistance corresponding to 29~T is denoted by red circles and the resistance at 56 T is denoted by red crosses. Note that the resistance for magnetic field higher than 29~T  was only measured for a reduced set of temperatures.}
\label{LSCO009e}
\end{figure}

In order to better explore the low temperature behaviour, and in particular to find out whether the sample is superconducting or resistive at low temperature, we measured a similar sample with the same Sr concentration (sample LSCO$_{0.09b}$ $x=0.09$) of a different origin and in a different high field facility (LNCMI) from room temperature down to 1.5~K.   This  sample probably has a slightly different Sr content from LSCO$_{009a}$, or is more disordered, which may explain the small differences in the resistance per square, and in the critical temperature. The plateau was retrieved for about 19~T and the corresponding square resistance is found to be around 5.8 k$\Omega $.  The measurements carried out  down to 1.5~K  at H$_C\simeq$19~T show that the system is indeed superconducting at T$\leq2$~K (See figure \ref{LSCO009e}). 
These measurements also seem to indicate that the actual insulator/superconductor transition at T=0 would be obtained for $H\simeq 37~T$ (about 2$H_C$) and R~$\simeq 11000~ \Omega$ (about 2$R_C$).

\section{Discussion}

In most of the data reported on QSIT in the literature, the resistance plateau is observed down to the lowest temperatures, presumably existing at zero temperature.  This observation constitutes a signature of the existence of a metallic state for a \textit{single} value of the tuning parameter.
However, Mason and coworkers \cite{Mason:2002} have shown experimentally that  a capacitively-coupled dissipative environment helps promote superconductivity at low temperature.  Also, Das and Doniach have inspected the possibility that pairs of vortices may first debind and then Bose-condense, which raises the possibility of the existence of a Bose metal between the superfluid and the Bose insulator, even at $T=0$ \cite{Das:1999,Das:2001}.  Such possibility of an intermediate metallic state has also been investigated by Galistki and coworkers \cite{Galitski:2005} using a two-fluid model.

We adopt here a different point of view, which takes into account the fact that our samples are not intrinsically 2D but rather lamellar and we investigate the consequences of a 2D-3D crossover.

\begin{figure}
\includegraphics[width=0.5\textwidth]{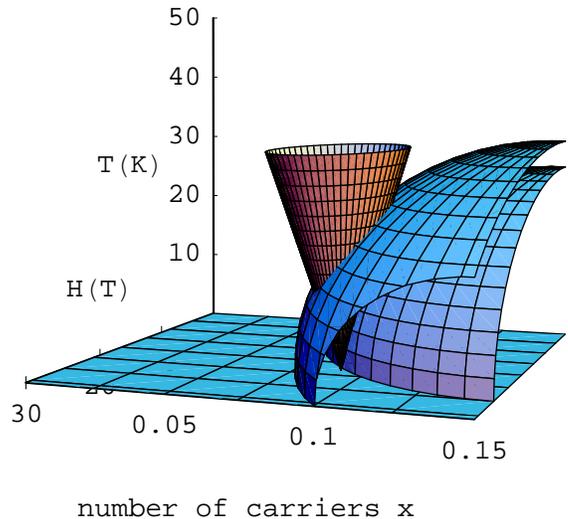}
\caption{(Color on line). Schematics of the 3D phase diagram in the (H,x,T) space. The lower dome represents the 2D critical temperature. The brown cone represents the domain of 2D critical fluctuations corresponding to a given value of (H$_C$, x$_C$), where the plateau is visible.  The second (upper) dome corresponds to the crossover temperature T$^{crossover}_{2D-3D}$ below which $\xi_C>$d. (See text).}
\label{phadia3D2}
\end{figure}

For this we suggest a three-dimensional (H,x,T) phase diagram displayed on Fig  \ref{phadia3D2}, that allows to picture the data presented in this paper as well as the data presented by Bollinger and coworkers \cite{Bollinger:2011}.
In this phase diagram, our measurements at a given T and x and as a function of H are represented by a line parallel to the H axis, whereas the measurements from Ref. \cite{Bollinger:2011} would be represented by a line parallel to the x axis in the H=0 plane. 
Bollinger et al. claim that a QCP is present at $x=0.06$, for H=0~T and T=0~K. They observe scaling laws around x= 0.06  with $\nu z =3/2$, when varying the number of carriers through varying the electrical field through the gate. An important difference with respect to our findings is that they observe the transition\textit{ in a single layer}. 
As a matter of facts, our sample with $x\simeq0.06$ does also present a plateau down to 2~K and for a rather low magnetic field (H$_C=1~T$). 
One has to bear in mind that our measurements are done at $H\neq0$ and that there is no such symmetry in magnetic field as in the electrical-field-driven case between $x<0.06$ and $x>0.06$, inasmuch as there is no way to go from insulating to superconducting by varying the magnetic field. Because of this, the "negative H$^2$" part of the phase diagram in figure \ref{phadia3D2} is unattainable. (Both H$>0$ and H$<0$ produce the same effect.)

Therefore in our experiment the range of investigation of the 3D phase diagram is extremely reduced  around  (H=0,x=0.06,T=0), whereas in the electrically driven data, the transition can be fully explored.

Our observations suggest one possible interpretation, related to the fact that, in contradiction to the system studied by Bollinger and coworkers, our systems are intrinsically lamellar.  When the coherence length is smaller than the interlayer spacing (between the CuO$_2$ planes) then the system can behave like a two-dimensional system in the Lawrence-Doniach sense \cite{Lawrence:1971}, but in a quantum way.  It may then be governed by a 2D quantum critical point.  However, when the temperature is decreased,  at some point, the coherence length gets larger than the interlayer spacing,  the system recovers classical 3D behaviour and the 2D QCP becomes irrelevant.  Then the effective "classical" 3D  transition takes place at much higher field (about 2H$_C$) and much higher square resistance (about 2R$_C$ for sample LSCO$_{0.09b}$).

Then the phase diagram would be similar to what  is represented in figure \ref{phadia3D2}. The pink-shaded dome represents the 2D phase transition.  Its intersection with the (T=0) plane is precisely the curve H$_C$(x), which is a line of QCP. For each point of the H$_C$(x) line at $T=0$,  a cone of quantum critical fluctuations may be drawn.  One of them is represented in brown in figure \ref{phadia3D2} for an arbitrary value of $(H_C (x),x)$.  Is also represented  another  arbitrary blue dome which mimics  the variation of the temperature $T_{2D-3D}^{crossover}$ below which $\xi_c\ge d$. The inside of this dome therefore corresponds to the 3D regime and is governed by another physics.

While cooling down the multilayer system exactly at (H$_C$(x),x), the representative point of the state of the system moves down along a vertical line inside a cone whose apex corresponds to $(H_C(x),x, T_C=0)$ and the resistance versus T exhibits a plateau. Therefore the system behaves like a 2D quantum critical material as long as the c-axis coherence length is smaller than the interlayer spacing. Whenever this remains true, the plateau is visible, but when approaching T=0, $\xi_C$ diverges and the representative point crosses the temperature below which $\xi_c\geq d$ (blue dome), then the system suddenly becomes 3D and the 2D-transition does not occur. Instead the system becomes superconducting with a finite 3D transition temperature. 
A test of this scenario would be to measure a single layer at different H$_C$ and x$_C$ values. In this case, the plateau should extend to the lowest measurable temperature. This would also allow to compare the critical exponents which seem to contrast when varying the electrical field ($\nu z =3/2$) or varying the magnetic field ($\nu z \sim 0.5-0.6$).

The variation of the critical square resistance at the plateau with doping (and therefore the non-universal value of the resistance) may also be related to this lamellar nature. If this universal value is related to self-duality as suggested by \cite{Fisher:1990b, Fisher:1990a}, the "symmetrical character" between the localization of vortices and of pairs may also be broken in a layered system since coupling from one layer to the other for pairs of electrons or for vortices is expected to be of different nature. Alternately, calculations in Ref  \cite{Galitski:2005} suggest that for more disordered systems, the transition will occur at higher resistance, which is consistent with our observations. 

\section{Conclusion}

We have investigated the magnetic field driven superconductor/insulator transition in eleven samples of LSCO thin films with various Sr content x, under high pulsed magnetic field. Our results indicate the presence of a plateau in the resistance versus temperature curves for every underdoped superconducting sample with $0.06 \leq x<0.1$, associated with scaling behavior with critical exponent $\nu z \simeq 0.5-0.6$ by contrast to electric field driven transition ($\nu z = 3/2$)  \cite{Bollinger:2011}. The critical field corresponding to the plateau is found to scale with the critical superconducting temperature with $1T\simeq1K$. However, these plateaus develop only in a limited range of temperatures, which may be reconciled with results from another group obtained on monolayers \cite{Bollinger:2011}, by taking into account the lamellar nature of our thin films. These results therefore point towards the existence of a 2D-quantum/3D crossover for the superconductor/insulator transition.

\section{Acknowledgement}
The authors gratefully acknowledges G. Scheerer for assistance during the pulsed field experiment at LNCMI and T. Wambeck during the pulsed field experiments at KU Leuven. B.L. also gratefully acknowledges fruitful discussions with M. Grilli and  S. Caprara. 
The work at the KU Leuven has been suported by the FWO Programmes and Methusalem Funding by the Flemish Government.  Research at ITT Kanpur has been supported by the J.C. Bose National Fellowship (R.C. Budhani). Part of this work has been founded by EuroMagNET II under the EU contract number 228043. B.L. acknowledges the ESF for support through the THIOX short visit grant n¡ 1081. This work has also been supported through a SESAME grant by the Ile-de-France Regional Counsel.

\bibliography{Mybib20}

\begin{thebibliography}{26}
\expandafter\ifx\csname natexlab\endcsname\relax\def\natexlab#1{#1}\fi
\expandafter\ifx\csname bibnamefont\endcsname\relax
  \def\bibnamefont#1{#1}\fi
\expandafter\ifx\csname bibfnamefont\endcsname\relax
  \def\bibfnamefont#1{#1}\fi
\expandafter\ifx\csname citenamefont\endcsname\relax
  \def\citenamefont#1{#1}\fi
\expandafter\ifx\csname url\endcsname\relax
  \def\url#1{\texttt{#1}}\fi
\expandafter\ifx\csname urlprefix\endcsname\relax\def\urlprefix{URL }\fi
\providecommand{\bibinfo}[2]{#2}
\providecommand{\eprint}[2][]{\url{#2}}

\bibitem[{\citenamefont{Lee and Ramakrishnan}(1985)}]{Lee:1985}
\bibinfo{author}{\bibfnamefont{P.}~\bibnamefont{Lee}} \bibnamefont{and}
  \bibinfo{author}{\bibfnamefont{T.}~\bibnamefont{Ramakrishnan}},
  \bibinfo{journal}{Rev. Mod. Phys.} \textbf{\bibinfo{volume}{57}},
  \bibinfo{pages}{287} (\bibinfo{year}{1985}).

\bibitem[{\citenamefont{Fisher}(1990)}]{Fisher:1990b}
\bibinfo{author}{\bibfnamefont{M.~P.~A.} \bibnamefont{Fisher}},
  \bibinfo{journal}{Phys. Rev. Letters} \textbf{\bibinfo{volume}{65}},
  \bibinfo{pages}{923} (\bibinfo{year}{1990}).

\bibitem[{\citenamefont{Fisher et~al.}(1990)\citenamefont{Fisher, Grinstein,
  and Girvin}}]{Fisher:1990a}
\bibinfo{author}{\bibfnamefont{M.~P.~A.} \bibnamefont{Fisher}},
  \bibinfo{author}{\bibfnamefont{G.}~\bibnamefont{Grinstein}},
  \bibnamefont{and} \bibinfo{author}{\bibfnamefont{S.}~\bibnamefont{Girvin}},
  \bibinfo{journal}{Phys. Rev. Letters} \textbf{\bibinfo{volume}{64}},
  \bibinfo{pages}{587} (\bibinfo{year}{1990}).

\bibitem[{\citenamefont{Galitski et~al.}(2005)\citenamefont{Galitski, Refael,
  Fisher, and Senthil}}]{Galitski:2005}
\bibinfo{author}{\bibfnamefont{V.~M.} \bibnamefont{Galitski}},
  \bibinfo{author}{\bibfnamefont{G.}~\bibnamefont{Refael}},
  \bibinfo{author}{\bibfnamefont{M.~P.~A.} \bibnamefont{Fisher}},
  \bibnamefont{and} \bibinfo{author}{\bibfnamefont{T.}~\bibnamefont{Senthil}},
  \bibinfo{journal}{Phys. Rev. Lett.} \textbf{\bibinfo{volume}{95}},
  \bibinfo{pages}{077002} (\bibinfo{year}{2005}),
  \urlprefix\url{http://link.aps.org/doi/10.1103/PhysRevLett.95.077002}.

\bibitem[{\citenamefont{Hebard and Paalanen}(1984)}]{Hebard:1984}
\bibinfo{author}{\bibfnamefont{A.}~\bibnamefont{Hebard}} \bibnamefont{and}
  \bibinfo{author}{\bibfnamefont{M.~A.} \bibnamefont{Paalanen}},
  \bibinfo{journal}{Phys. Rev. B} \textbf{\bibinfo{volume}{R30}},
  \bibinfo{pages}{4063} (\bibinfo{year}{1984}).

\bibitem[{\citenamefont{Hebard and Paalanen}(1990)}]{Hebard:1990}
\bibinfo{author}{\bibfnamefont{A.}~\bibnamefont{Hebard}} \bibnamefont{and}
  \bibinfo{author}{\bibfnamefont{M.~A.} \bibnamefont{Paalanen}},
  \bibinfo{journal}{Phys. Rev. Letters} \textbf{\bibinfo{volume}{65}},
  \bibinfo{pages}{927} (\bibinfo{year}{1990}).

\bibitem[{\citenamefont{Haviland et~al.}(1989)\citenamefont{Haviland, Liu, and
  Goldman}}]{Haviland:1989}
\bibinfo{author}{\bibfnamefont{D.}~\bibnamefont{Haviland}},
  \bibinfo{author}{\bibfnamefont{Y.}~\bibnamefont{Liu}}, \bibnamefont{and}
  \bibinfo{author}{\bibfnamefont{A.~M.} \bibnamefont{Goldman}},
  \bibinfo{journal}{Phys. Rev. Lett.} \textbf{\bibinfo{volume}{62}},
  \bibinfo{pages}{2181} (\bibinfo{year}{1989}).

\bibitem[{\citenamefont{Graybeal and Beasley}(1984)}]{Graybeal:1984}
\bibinfo{author}{\bibfnamefont{J.}~\bibnamefont{Graybeal}} \bibnamefont{and}
  \bibinfo{author}{\bibfnamefont{M.}~\bibnamefont{Beasley}},
  \bibinfo{journal}{Phys. Rev. B} \textbf{\bibinfo{volume}{29}},
  \bibinfo{pages}{4167} (\bibinfo{year}{1984}).

\bibitem[{\citenamefont{Yazdani and Kapitulnik}(1995)}]{Yazdani:1995}
\bibinfo{author}{\bibfnamefont{A.}~\bibnamefont{Yazdani}} \bibnamefont{and}
  \bibinfo{author}{\bibfnamefont{A.}~\bibnamefont{Kapitulnik}},
  \bibinfo{journal}{Phys. Rev. Lett.} \textbf{\bibinfo{volume}{74}},
  \bibinfo{pages}{3037} (\bibinfo{year}{1995}).

\bibitem[{\citenamefont{Aubin et~al.}(2006)\citenamefont{Aubin,
  Marrache-Kikuchi, Pourret, Behnia, BergŽ, Dumoulin, and
  Lesueur}}]{Aubin:2006}
\bibinfo{author}{\bibfnamefont{H.}~\bibnamefont{Aubin}},
  \bibinfo{author}{\bibfnamefont{C.~A.} \bibnamefont{Marrache-Kikuchi}},
  \bibinfo{author}{\bibnamefont{Pourret}},
  \bibinfo{author}{\bibfnamefont{K.}~\bibnamefont{Behnia}},
  \bibinfo{author}{\bibfnamefont{L.}~\bibnamefont{BergŽ}},
  \bibinfo{author}{\bibfnamefont{L.}~\bibnamefont{Dumoulin}}, \bibnamefont{and}
  \bibinfo{author}{\bibfnamefont{J.}~\bibnamefont{Lesueur}},
  \bibinfo{journal}{Phys. Rev. B} \textbf{\bibinfo{volume}{73}},
  \bibinfo{pages}{094521} (\bibinfo{year}{2006}).

\bibitem[{\citenamefont{Steiner et~al.}(2005)\citenamefont{Steiner, Boebinger,
  and Kapitulnik}}]{Steiner:2005}
\bibinfo{author}{\bibfnamefont{M.~A.} \bibnamefont{Steiner}},
  \bibinfo{author}{\bibfnamefont{G.}~\bibnamefont{Boebinger}},
  \bibnamefont{and}
  \bibinfo{author}{\bibfnamefont{A.}~\bibnamefont{Kapitulnik}},
  \bibinfo{journal}{Phys. Rev. Letters} \textbf{\bibinfo{volume}{94}},
  \bibinfo{pages}{107008} (\bibinfo{year}{2005}).

\bibitem[{\citenamefont{Wang et~al.}(1991)\citenamefont{Wang, Beauchamp,
  Berkley, Johnson, Liu, Zhang, and Goldman}}]{Wang:1991}
\bibinfo{author}{\bibfnamefont{T.}~\bibnamefont{Wang}},
  \bibinfo{author}{\bibfnamefont{K.~M.} \bibnamefont{Beauchamp}},
  \bibinfo{author}{\bibfnamefont{D.~D.} \bibnamefont{Berkley}},
  \bibinfo{author}{\bibfnamefont{B.~R.} \bibnamefont{Johnson}},
  \bibinfo{author}{\bibfnamefont{J.-X.} \bibnamefont{Liu}},
  \bibinfo{author}{\bibfnamefont{J.}~\bibnamefont{Zhang}}, \bibnamefont{and}
  \bibinfo{author}{\bibfnamefont{A.~M.} \bibnamefont{Goldman}},
  \bibinfo{journal}{Phys. Rev. B} \textbf{\bibinfo{volume}{43}},
  \bibinfo{pages}{8623} (\bibinfo{year}{1991}),
  \urlprefix\url{http://link.aps.org/doi/10.1103/PhysRevB.43.8623}.

\bibitem[{\citenamefont{Bollinger et~al.}(2011)\citenamefont{Bollinger, Dubuis,
  Yoon, Pavuna, Misewich, and Bozÿovic«}}]{Bollinger:2011}
\bibinfo{author}{\bibfnamefont{A.~T.} \bibnamefont{Bollinger}},
  \bibinfo{author}{\bibfnamefont{G.}~\bibnamefont{Dubuis}},
  \bibinfo{author}{\bibfnamefont{J.}~\bibnamefont{Yoon}},
  \bibinfo{author}{\bibfnamefont{D.}~\bibnamefont{Pavuna}},
  \bibinfo{author}{\bibfnamefont{J.}~\bibnamefont{Misewich}}, \bibnamefont{and}
  \bibinfo{author}{\bibfnamefont{I.}~\bibnamefont{Bozÿovic«}},
  \bibinfo{journal}{Nature} \textbf{\bibinfo{volume}{472}},
  \bibinfo{pages}{458} (\bibinfo{year}{2011}).

\bibitem[{\citenamefont{Leng et~al.}(2011)\citenamefont{Leng,
  Garcia-Barriocanal, Bose, Lee, and Goldman}}]{Leng:2011}
\bibinfo{author}{\bibfnamefont{X.}~\bibnamefont{Leng}},
  \bibinfo{author}{\bibfnamefont{J.}~\bibnamefont{Garcia-Barriocanal}},
  \bibinfo{author}{\bibfnamefont{S.}~\bibnamefont{Bose}},
  \bibinfo{author}{\bibfnamefont{Y.}~\bibnamefont{Lee}}, \bibnamefont{and}
  \bibinfo{author}{\bibfnamefont{A.~M.} \bibnamefont{Goldman}},
  \bibinfo{journal}{Phys. Rev. Lett.} \textbf{\bibinfo{volume}{107}}
  (\bibinfo{year}{2011}).

\bibitem[{\citenamefont{P.Wagner et~al.}(2001)\citenamefont{P.Wagner, Ruan,
  Gordon, Vanacken, Moshchalkov, and Bruynseraede}}]{Wagner:2001}
\bibinfo{author}{\bibnamefont{P.Wagner}},
  \bibinfo{author}{\bibfnamefont{K.}~\bibnamefont{Ruan}},
  \bibinfo{author}{\bibfnamefont{I.}~\bibnamefont{Gordon}},
  \bibinfo{author}{\bibfnamefont{J.}~\bibnamefont{Vanacken}},
  \bibinfo{author}{\bibfnamefont{V.~V.} \bibnamefont{Moshchalkov}},
  \bibnamefont{and}
  \bibinfo{author}{\bibfnamefont{Y.}~\bibnamefont{Bruynseraede}},
  \bibinfo{journal}{Physica C} \textbf{\bibinfo{volume}{356}},
  \bibinfo{pages}{107} (\bibinfo{year}{2001}).

\bibitem[{\citenamefont{Rout and Budhani}(2010)}]{Rout:2010}
\bibinfo{author}{\bibfnamefont{P.~K.} \bibnamefont{Rout}} \bibnamefont{and}
  \bibinfo{author}{\bibfnamefont{R.~C.} \bibnamefont{Budhani}},
  \bibinfo{journal}{Phys. Rev. B} \textbf{\bibinfo{volume}{82}},
  \bibinfo{pages}{024518} (\bibinfo{year}{2010}),
  \urlprefix\url{http://link.aps.org/doi/10.1103/PhysRevB.82.024518}.

\bibitem[{\citenamefont{Ando et~al.}(1995)\citenamefont{Ando, Boebinger,
  Passner, Kimura, and Kishio}}]{Ando:1995}
\bibinfo{author}{\bibfnamefont{Y.}~\bibnamefont{Ando}},
  \bibinfo{author}{\bibfnamefont{G.}~\bibnamefont{Boebinger}},
  \bibinfo{author}{\bibfnamefont{A.}~\bibnamefont{Passner}},
  \bibinfo{author}{\bibfnamefont{T.}~\bibnamefont{Kimura}}, \bibnamefont{and}
  \bibinfo{author}{\bibfnamefont{K.}~\bibnamefont{Kishio}},
  \bibinfo{journal}{Phys. Rev. Letters} \textbf{\bibinfo{volume}{75}},
  \bibinfo{pages}{25} (\bibinfo{year}{1995}).

\bibitem[{\citenamefont{Leridon et~al.}(2007)\citenamefont{Leridon, Vanacken,
  Wambecq, and Moshchalkov}}]{Leridon:2007b}
\bibinfo{author}{\bibfnamefont{B.}~\bibnamefont{Leridon}},
  \bibinfo{author}{\bibfnamefont{J.}~\bibnamefont{Vanacken}},
  \bibinfo{author}{\bibfnamefont{T.}~\bibnamefont{Wambecq}}, \bibnamefont{and}
  \bibinfo{author}{\bibfnamefont{V.~V.} \bibnamefont{Moshchalkov}},
  \bibinfo{journal}{Phys. Rev. B} \textbf{\bibinfo{volume}{76}},
  \bibinfo{pages}{012503} (\bibinfo{year}{2007}).

\bibitem[{\citenamefont{Das and Doniach}(1999)}]{Das:1999}
\bibinfo{author}{\bibfnamefont{D.}~\bibnamefont{Das}} \bibnamefont{and}
  \bibinfo{author}{\bibfnamefont{S.}~\bibnamefont{Doniach}},
  \bibinfo{journal}{Phys. Rev. Letters} \textbf{\bibinfo{volume}{60}},
  \bibinfo{pages}{1261} (\bibinfo{year}{1999}).

\bibitem[{\citenamefont{Das and Doniach}(2001)}]{Das:2001}
\bibinfo{author}{\bibfnamefont{D.}~\bibnamefont{Das}} \bibnamefont{and}
  \bibinfo{author}{\bibfnamefont{S.}~\bibnamefont{Doniach}},
  \bibinfo{journal}{Phys. Rev. Letters} \textbf{\bibinfo{volume}{64}},
  \bibinfo{pages}{134511} (\bibinfo{year}{2001}).

\bibitem[{\citenamefont{Weckhuysen}(2002)}]{Weckuysen:2002b}
\bibinfo{author}{\bibfnamefont{L.}~\bibnamefont{Weckhuysen}}
  (\bibinfo{year}{2002}).

\bibitem[{\citenamefont{Deutscher et~al.}(1980)\citenamefont{Deutscher,
  Bandyopadhyay, Chui, Lindenfeld, McLean, and Worthington}}]{Deutscher:1980}
\bibinfo{author}{\bibfnamefont{G.}~\bibnamefont{Deutscher}},
  \bibinfo{author}{\bibfnamefont{B.}~\bibnamefont{Bandyopadhyay}},
  \bibinfo{author}{\bibfnamefont{T.}~\bibnamefont{Chui}},
  \bibinfo{author}{\bibfnamefont{P.}~\bibnamefont{Lindenfeld}},
  \bibinfo{author}{\bibfnamefont{W.}~\bibnamefont{McLean}}, \bibnamefont{and}
  \bibinfo{author}{\bibfnamefont{T.}~\bibnamefont{Worthington}},
  \bibinfo{journal}{Phys. Rev. Letters} \textbf{\bibinfo{volume}{44}},
  \bibinfo{pages}{1150} (\bibinfo{year}{1980}).

\bibitem[{\citenamefont{Simon et~al.}(1987)\citenamefont{Simon, Dalrymple,
  Vechten, Fuller, and Wolf}}]{Simon:1987}
\bibinfo{author}{\bibfnamefont{R.}~\bibnamefont{Simon}},
  \bibinfo{author}{\bibfnamefont{B.~J.} \bibnamefont{Dalrymple}},
  \bibinfo{author}{\bibfnamefont{D.~V.} \bibnamefont{Vechten}},
  \bibinfo{author}{\bibfnamefont{W.}~\bibnamefont{Fuller}}, \bibnamefont{and}
  \bibinfo{author}{\bibfnamefont{S.}~\bibnamefont{Wolf}},
  \bibinfo{journal}{Phys. Rev. B} \textbf{\bibinfo{volume}{36}},
  \bibinfo{pages}{1962} (\bibinfo{year}{1987}).

\bibitem[{\citenamefont{Parendo et~al.}(2005)\citenamefont{Parendo, Tan,
  Bhattacharya, Eblen-Zayas, Staley, and Goldman}}]{Parendo:2005}
\bibinfo{author}{\bibfnamefont{K.~A.} \bibnamefont{Parendo}},
  \bibinfo{author}{\bibfnamefont{K.~H. S.~B.} \bibnamefont{Tan}},
  \bibinfo{author}{\bibfnamefont{A.}~\bibnamefont{Bhattacharya}},
  \bibinfo{author}{\bibfnamefont{M.}~\bibnamefont{Eblen-Zayas}},
  \bibinfo{author}{\bibfnamefont{N.~E.} \bibnamefont{Staley}},
  \bibnamefont{and} \bibinfo{author}{\bibfnamefont{A.~M.}
  \bibnamefont{Goldman}}, \bibinfo{journal}{Phys. Rev. Lett.}
  \textbf{\bibinfo{volume}{94}}, \bibinfo{pages}{197004}
  (\bibinfo{year}{2005}),
  \urlprefix\url{http://link.aps.org/doi/10.1103/PhysRevLett.94.197004}.

\bibitem[{\citenamefont{Mason and Kapitulnik}(2002)}]{Mason:2002}
\bibinfo{author}{\bibfnamefont{N.}~\bibnamefont{Mason}} \bibnamefont{and}
  \bibinfo{author}{\bibfnamefont{A.}~\bibnamefont{Kapitulnik}},
  \bibinfo{journal}{Phys. Rev. B} \textbf{\bibinfo{volume}{65}},
  \bibinfo{pages}{220505R} (\bibinfo{year}{2002}).

\bibitem[{\citenamefont{Lawrence and Doniach}(1971)}]{Lawrence:1971}
\bibinfo{author}{\bibfnamefont{W.}~\bibnamefont{Lawrence}} \bibnamefont{and}
  \bibinfo{author}{\bibfnamefont{S.}~\bibnamefont{Doniach}},
  \bibinfo{journal}{Proc. 12th Int. Conf. on Low Temp. Phys.} p.
  \bibinfo{pages}{361} (\bibinfo{year}{1971}).

\end{thebibliography}

\end{document}